\documentclass{cmspaper}
\usepackage{epsfig}
\usepackage{amsmath}
\usepackage{here}
\begin{document}

\renewcommand{\arraystretch}{1.5}

\begin{titlepage}

\flushright{\bf\Large IEKP-KA/2001-28}

   \cmsnote{2002/006}
   \date{19 January 2002}

  \boldmath
  \title{Prospects for Higgs Boson Searches in the Channel $W^\pm H^0 \rightarrow l^\pm \nu b\bar{b}$}
  \unboldmath

  \begin{Authlist}
    V.~Drollinger\Aref{a} and Th.~M\"uller
       \Instfoot{iekp}{IEKP, Karlsruhe University, Germany}
    D.~Denegri
       \Instfoot{cern}{CERN. Geneva, Switzerland and DAPNIA Saclay, France}
  \end{Authlist}
  \Anotfoot{a}{Now at Department of Physics and Astronomy, University of New Mexico, USA}


  \begin{abstract}

We present a method how to detect the $W^\pm H^0 \rightarrow l^\pm \nu b\bar{b}$ in the high luminosity LHC environment with the CMS detector. This study is performed with fast detector response simulation including high luminosity event pile up. The main aspects of reconstruction are pile up jet rejection, identification of $b$-jets and improvement of Higgs mass resolution.

The detection potential in the SM for $m_{H^0} \le$ 130 $GeV/c^2$ and in the MSSM is only encouraging for high integrated luminosity. Nevertheless it is possible to extract important Higgs parameters which are useful to elucidate the nature of the Higgs sector. In combination with other channels, this channel provides valuable information on Higgs boson couplings.

  \end{abstract} 

  
\end{titlepage}

\section{Introduction}

\begin{figure}[ht]
\begin{center}
 \includegraphics[width=0.42\textwidth,angle=+0]{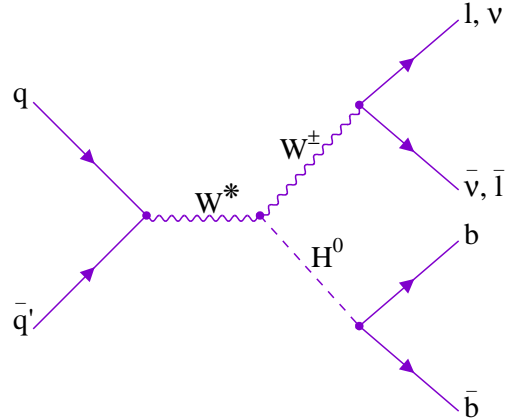}
 \caption{\sl $W^\pm H^0 \rightarrow l^\pm \nu b\bar{b}$ signal event at LO.\rm}
 \label{fig:fey_wh}
\end{center}
\end{figure}

The observation of at least one Higgs boson is an important proof of the Higgs mechanism \cite{HiggsMech} which is introduced to explain the masses of elementary particles. Beside the discovery of a Higgs boson, it is also important to study the Higgs boson couplings. Higgs bosons lighter than 130 $GeV/c^2$ decay mainly to a $b\bar{b}$ pair \cite{HDECAY}. In this note we describe a method to observe a Higgs boson in the associated production channel $W^\pm H^0 \rightarrow l^\pm \nu b\bar{b}$, shown in Figure~\ref{fig:fey_wh}, with leptonically decaying $W^\pm$. Among other production channels with $H^0 \rightarrow b\bar{b}$ decay \cite{THESIS}, the $W^\pm H^0$ channel turns out to have a low signal to background ratio and therefore requires a large integrated luminosity.
The signal and backgrounds are simulated in the high luminosity regime with each event superimposed on an average of 17.3 Poisson distributed pile up events generated with PYTHIA \cite{PYTHIA} (MSTP(2) = 2 , MSTP(33) = 3 , MSTP(81) = 1 , MSTP(82) = 4 and PARP(82) = 3.). The relevant signal and background cross sections at the LHC ($\sqrt{s_{pp}} = $ 14 $TeV$) and particle masses used in the simulation are listed in Table~\ref{tab:crossections}.

\vspace*{5mm}
\begin{table}[ht]
 \begin{center}
 \begin{tabular}{|lcr|lcr|}
 \hline
 \multicolumn{3}{|c|}{LO cross sections} & \multicolumn{3}{|c|}{masses}\\
 \hline
  $\sigma_{W^\pm H^0_{SM}} \times BR_{H^0_{SM} \rightarrow b\bar{b}}$ & = 
 & 2.51 - 0.41 $pb$ &             $m_{H^0_{SM}}$ & = & 90 - 135 $GeV/c^2$ \\
 $\sigma_{W^\pm Z^0}$ & = & 27   $pb$ &   $m_{Z^0}$ & = & 91.187 $GeV/c^2$\\
  $\sigma_{W^\pm jj}$ & = & 30   $nb$ & $m_{W^\pm}$ & = & 80.41  $GeV/c^2$\\
  $\sigma_{t\bar{t}}$ & = & 570  $pb$ &     $m_{t}$ & = & 173.8  $GeV/c^2$\\
  $\sigma_{t\bar{b}}$ & = & 320  $pb$ &     $m_{b}$ & = & 4.3    $GeV/c^2$\\
 \hline
 \end{tabular}
 \end{center}
  \caption{\sl PYTHIA cross sections of signal and background relevant for the $W^\pm H^0 \rightarrow l^\pm \nu b\bar{b}$ channel and calculated with parton density function CTEQ4l \cite{PDFLIB}. The branching ratio of the $W^\pm$ decay to electrons or muons (= 22\%) is not included in the cross sections of this table. The particle masses are from \cite{PDG98}. The generation of single top ($t\bar{b}$) events can be done in PYTHIA with processes number 2 and 83 by forcing $\sqrt{\hat{s}} >$ 180 $GeV$ \cite{hartmut}.\rm}
\label{tab:crossections}
\end{table}
After the event generation, the detector simulation is performed: FATSIM \cite{FATSIM} is used for the simulation of tracker response and track reconstruction. Track momentum smearing, impact parameter smearing, impact parameter tails and track reconstruction efficiency are taken into account, as well as geometrical acceptances. CMSJET \cite{CMSJET} is used for the simulation of calorimeter response, jet reconstruction, missing transverse energy calculation and lepton smearing. The reconstruction efficiency for leptons (in this study only electrons and muons) is assumed to be 90\%. All parametrisations have been obtained from GEANT based detailed simulations.


\section{Reconstruction}

\begin{figure}[ht]
\begin{center}
 \includegraphics[width=0.62\textwidth,angle=+0]{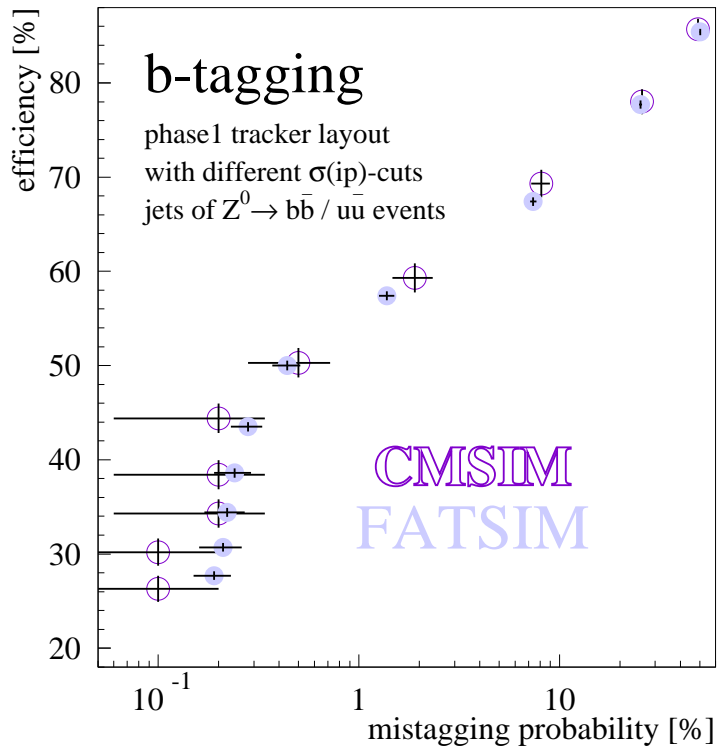}
 \caption{\sl $b$-tagging performances of $Z^0 \rightarrow jj$ events by FATSIM
          (bright) and CMSIM (dark). The tagging rates of $b$- and $u$-jets were determined with the following algorithm: 2 tracks in a cone of 0.4 around the jet ($E_T > $ 20~GeV, $|\eta| < $ 2.4) axis with $p_T > $ 0.9~GeV/c, min.~6 hits including 2 pixel hits (in the case of CMSIM) , $ip <  2$~mm and the signed transverse impact parameter significance $\sigma(ip) > $ 0.0 ... 4.5 (various cuts).}
 \label{fig:ztag_p1}
\end{center}
\end{figure}

A typical signal event as shown in Figure~\ref{fig:fey_wh} is expected to give a final state consisting of one isolated lepton, missing transverse energy $E_T^{miss}$ and two $b$-jets. Pile up events, underlying event and gluon radiation are sources of additional jets which complicate the reconstruction and selection. An initial study \cite{WH_DIP} of this channel showed already the importance of two reconstruction features, $b$-tagging performance and jet reconstruction:

A clean and efficient identification of $b$-jets is important to reduce most of the background with less than two genuine $b$-jets. For a realistic (and reasonably fast) $b$-tagging simulation, the understanding of impact parameters, impact parameter errors and track reconstruction efficiency are crucial. In Figure~\ref{fig:ztag_p1} the parametrisation FATSIM (fast tracker simulation) is compared with the detailed simulation: the $b$-tagging performance of the CMS tracker (phase 1) is simulated for soft jets coming from the $Z^0$ decay using mainly the significance of the signed transverse impact parameter $\sigma(ip)$ cuts on two tracks per jet. There is a good agreement between the two simulations. In this example, the $b$-tagging efficiency for $\sigma(ip) > $ 2 is 50\% and the mistagging probability is less than 1\%. The $b$-tagging efficiency for jets from the Higgs decay is higher ($\approx$ 60\%), because the jets coming from the heavier Higgs boson are more energetically. Identification of leptons inside jets and reconstruction of secondary vertices can improve the $b$-tagging performance further.
\vspace*{5mm}

A good Higgs mass resolution helps to enhance the signal event invariant $m(j,j)$ mass peak allowing a better visibility of the signal and a more precise determination of the Higgs mass. A good mass resolution can be obtained, when the energy and direction of each reconstructed jet agree as closely as possible with the quantities of the corresponding parent quark. This can be achieved with jet corrections as described in \cite{JETCOR,JETRAD}: after jets are reconstructed with the UA1 cone ($R_C = $ 0.4) algorithm, their quality is improved in two steps. Firstly, single jets are corrected by taking into account out of cone energy and extra energy which does not come from the original parton. This energy correction improves the Higgs mass resolution by 35\%. In the second step, the final state gluon radiation is corrected (FSR correction). Herefore two jets are combined into one jet taking into account geometrical effects. The second step gives an additional 10\% improvement. Clearly, the single jet corrections are more important than FSR corrections, but nevertheless the FSR corrections can still improve the Higgs mass resolution significantly. Further improvements, not studied here, are possible by using tracks.

A detailed description of the analysis follows:

\noindent
{\bf\boldmath$\diamond$ Trigger and Preselection}\\
Events are triggered, if there is one isolated lepton with $p_T >$ 20 $GeV/c$ in the tracker acceptance. A lepton is considered as isolated, if there are no additional tracks with $p_T >$ 2 $GeV/c$ in a cone of 0.3 around the lepton. The lepton reconstruction efficiency is assumed to be 90\%. The preselection includes the requirement of at least two jets with $E_T >$ 30 $GeV$ and $|\eta| <$ 2.5.

\begin{figure}[ht]
\begin{center}
 \includegraphics[width=0.80\textwidth,angle=+0]{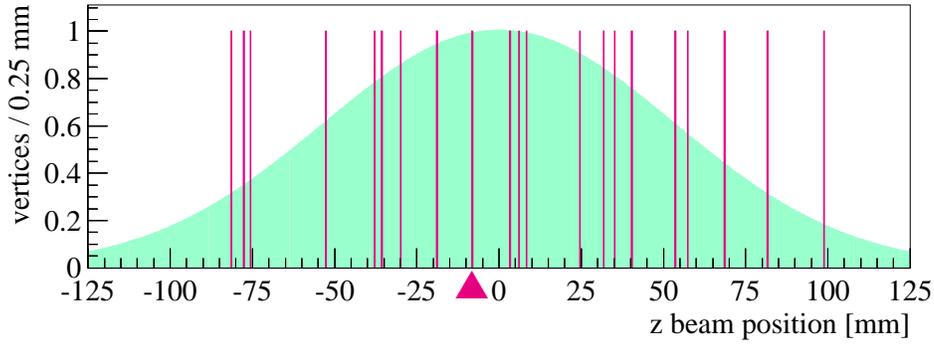}
 \caption{\sl Example of bunch crossing spread in z direction: 21 spikes indicate the primary $pp$ collision vertices due to one hard interaction (marked with a triangle) and twenty superimposed pile up events. The bin width of 250 $\mu m$ corresponds to ten times the resolution of the z coordinate of the primary vertex $V^0_Z$. The shaded area is a measure of the expected frequency distribution of primary vertices.}
 \label{fig:v0_beam_example}
\end{center}
\end{figure}

\noindent
{\bf\boldmath$\diamond$ Pile Up Jet Rejection}\\
In order to be able to count the jets of one particular event, pile up jets are removed from the event: jets are removed, if they contain no track with $p_T >$ 2 $GeV/c$ in a cone of 0.4 around the jet axis, or if they contain a track of $p_T >$ 2 $GeV/c$ in a cone of 0.2 around the jet axis which is coming from the wrong primary vertex: $\Delta(V^{track}_Z,V^0_Z) <$ 250 $\mu m$. The identification of primary vertices is possible, because the isolated lepton points to the correct primary vertex and the primary vertex z coordinate ($V^0_Z$) resolution is as good as 25 $\mu m$, whereas the z bunch crossing spread has $\sigma =$~53~$mm$, as illustrated in Figure~\ref{fig:v0_beam_example}.
The performance of this procedure depends basically on instantaneous luminosity (number of pile up events) and $V^0_Z$ resolution of the CMS tracker.
\vspace*{3mm}

\noindent
{\bf\boldmath$\diamond$ $b$-tagging}\\
Two tagged jets are required per event. A jet is considered as $b$-tagged, if there are two tracks with impact parameter significance $\sigma(ip) > $ 2.0, or if there are three tracks with $\sigma(ip) > $ 1.6, or if there are two tracks with $\sigma(ip) > $ 1.0 and a lepton inside the jet. Here $ip$ is the signed transverse impact parameter of a track with $p_T >$ 0.9 $GeV/c$ and $ip < $ 0.2 $cm$.

\noindent
{\bf\boldmath$\diamond$ Jet Corrections}\\
The tagged jets are corrected as $b$-jets with or without lepton and additional jets which are closer than 0.6 to one of these jets are corrected as gluon jets and combined with the $b$-jets according to the procedure described in \cite{JETCOR,JETRAD}.

\noindent
{\bf\boldmath$\diamond$ Jet and Lepton Veto}\\
If there is an additional jet with $E_T >$ 20 $GeV$ and $|\eta| <$ 2.5, the event is rejected. Also events with an additional isolated lepton ($p_T >$ 5 $GeV/c$) are not accepted. The jet veto is only possible after ``pile up jet rejection''.

\noindent
{\bf\boldmath$\diamond$ Relative $E_T$ Balance}\\
The relative $E_T$ balance is defined as the ratio of the transverse energy of the event ($E_T$ of $b,\bar{b},l$ and $\nu$ which is ideally close to zero) and the $E_T$ sum of the four objects expected in the event.
\begin{align}
 \frac{E_T(b,\bar{b},l,\nu)}{E_T(b)+E_T(\bar{b})+E_T(l)+E_T(\nu)} < 0.15 \qquad E_T(\nu) \cong E_T^{miss} \notag
\end{align}

\noindent
{\bf\boldmath$\diamond$ $W^\pm$ Identification}
The reconstruction of the transverse mass of the W boson is intended to suppress further badly reconstructed events. It turns out that only an upper cut improves the signal visibility.
\begin{align}
m_T(l,\nu) <  100\ GeV/c^2 \notag
\end{align}

\noindent
{\bf\boldmath$\diamond$ Higgs Mass Reconstruction}
The invariant mass of both fully corrected $b$-jets is calculated. Finally events are counted which satisfy $\bar{m}\ \pm$ 1.4 $\sigma$ with $\bar{m}$ and $\sigma$ from Table~\ref{tab:wh-ma-wi}.
\vspace*{5mm}
\begin{table}[ht]
 \begin{center}
 \begin{tabular}{|cccccccccccc|}
 \hline
 $m_{H^0}$ &=&  90 &  95 & 100 & 105 & 110 & 115 & 120 & 125 & 130 & 135 \\
 $\bar{m}$ &=&91.28&95.67&100.1&104.5&108.5&113.5&117.8&122.0&126.5&131.2\\
 $\sigma $ &=&10.76&11.04&11.19&11.90&11.67&11.66&12.28&12.68&13.22&13.73\\
 \hline
 \end{tabular}
 \end{center}
  \caption[Reconstructed Masses an Widths]{\sl Generated $m_{H^0}$, reconstructed $\bar{m}$ and $\sigma$ in units of $GeV/c^2$. $m_{inv}(j,j)$ of the pure signal is fitted with a Gaussian to obtain the masses and widths used in $\bar{m}\ \pm$ 1.4 $\sigma$.\rm}
\label{tab:wh-ma-wi}
\end{table}

The main background rejection comes from $b$-tagging. The non-$b$-jet backgrounds (mainly $W^\pm jj$) are reduced strongly, but backgrounds with two genuine $b$-jets are not suppressed. A large fraction of the top background can be removed with the combination of ``jet and lepton veto'', ``relative $E_T$ balance'' and ``$W^\pm$ identification''. These steps also help to remove events with a large amount of unclustered energy and badly reconstructed events. All backgrounds are reduced by the mass window cut which is more effective with improved Higgs mass resolution.
\vspace*{5mm}
\begin{figure}[ht]
\begin{center}
 \includegraphics[width=0.57\textwidth,angle=+0]{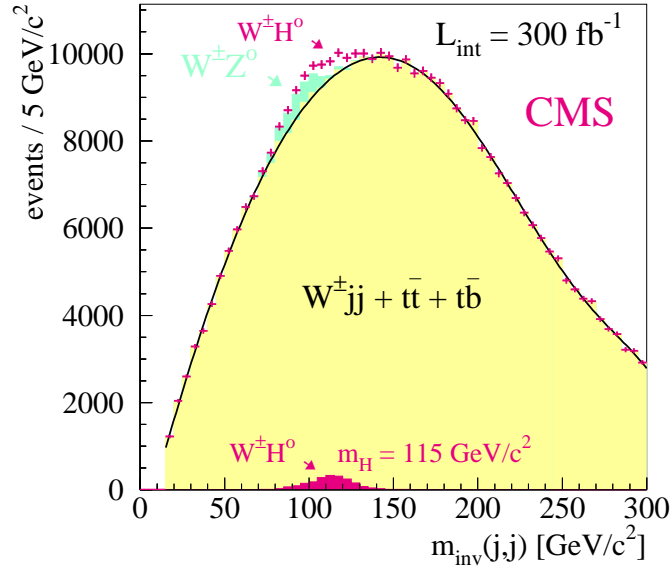}
 \caption{\sl $W^\pm H^0$ signal (white or dark shaded) plus resonant $W^\pm Z^0$ background (shaded) plus non resonant background (light shaded). The crosses are signal plus background with statistical error bars and are well above the background in the signal region. Simulated Higgs mass is $m_{H^0} =$ 115 $GeV/c^2$ with $L_{int} = $ 300 $fb^{-1}$.\rm}
 \label{fig:wh_sig_bg}
\end{center}
\end{figure}

\begin{figure}[ht]
\begin{center}
 \includegraphics[width=0.49\textwidth,angle=+0]{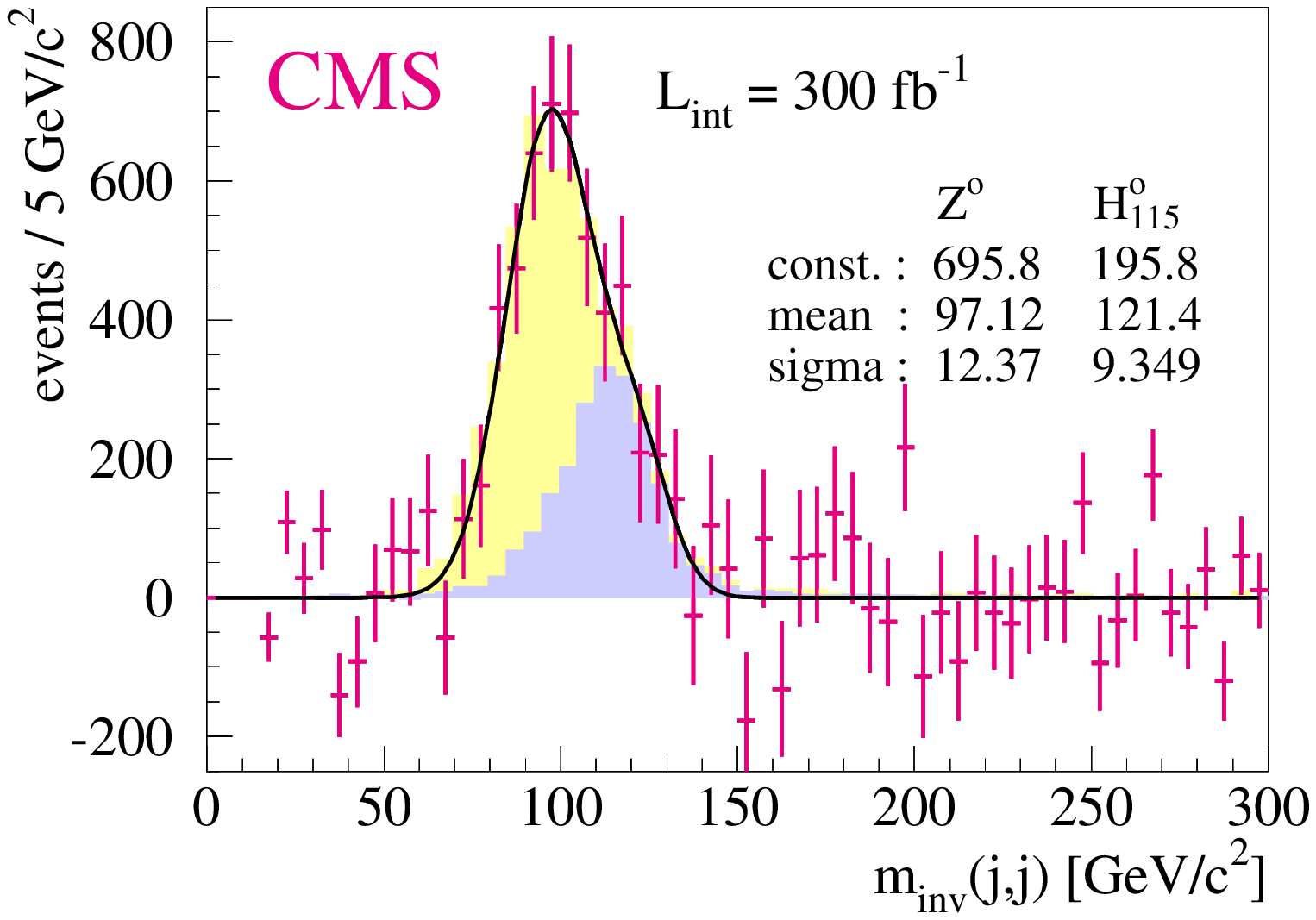}
 \includegraphics[width=0.49\textwidth,angle=+0]{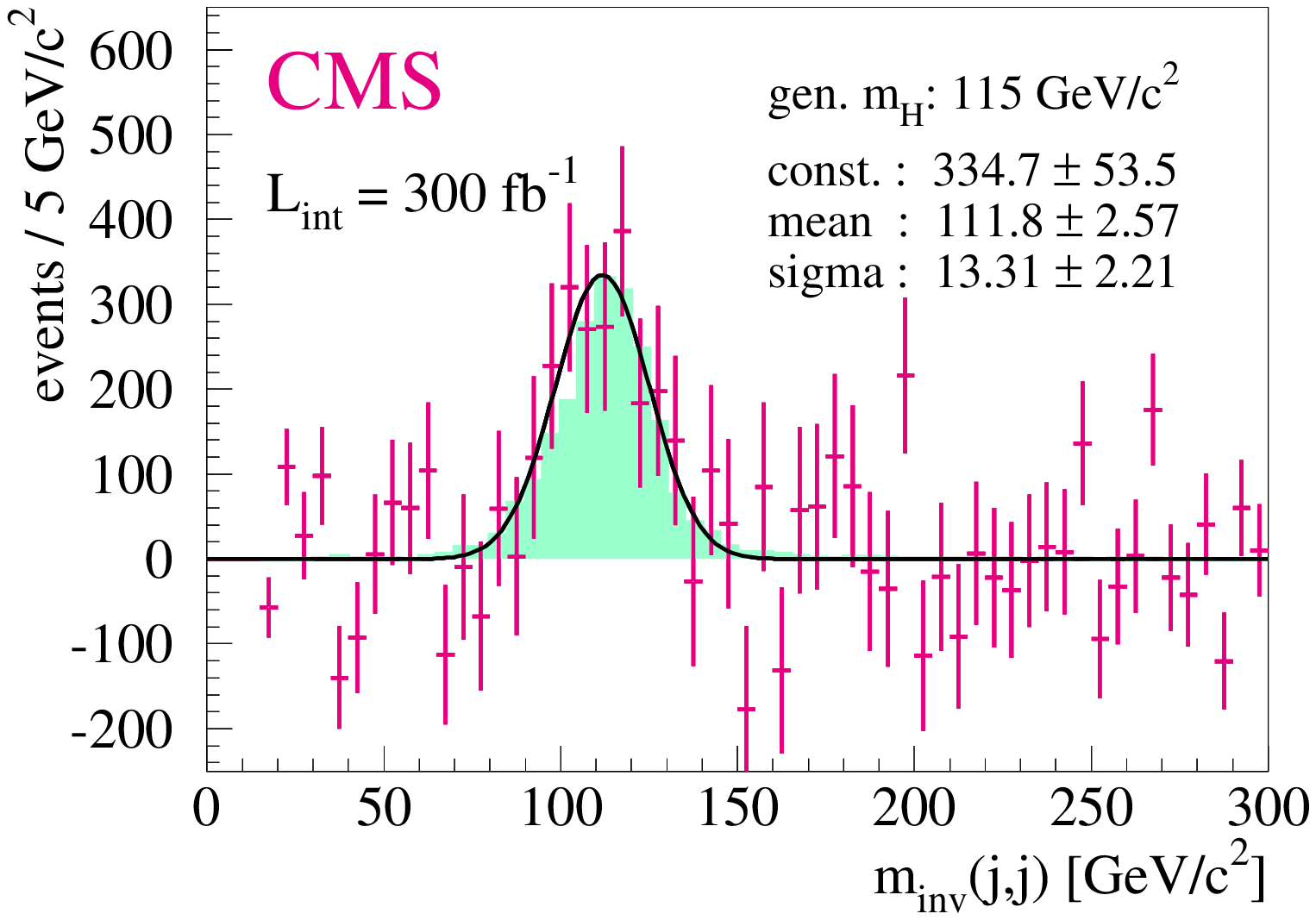}
 \caption{\sl Left: signal plus background after subtraction of non resonant background. The simulated data points are fitted with two Gaussians. Right: signal plus background after subtraction of all backgrounds. The simulated data points are fitted with a Gaussian which describes the pure $W^\pm H^0$ signal (shaded) well.
\rm}
 \label{fig:wh_peaks}
\end{center}
\end{figure}

For an integrated luminosity of 300 $fb^{-1}$, 1610 $W^\pm H^0$ events ($m_{H^0} =$ 115 $GeV/c^2$) are selected with an efficiency of 5\%, starting from all triggered events. 1198 $W^\pm Z^0$ events are selected with an efficiency of 0.3\%. The efficiency for $W^\pm jj$ events (27565) is 0.003\%, for $t\bar{t}$ events (36089) is 0.1\% and for $t\bar{b}$ events (6096) is 0.09\%. Figure~\ref{fig:wh_sig_bg} shows the expected signal plus background. The signal peak can be hardly seen. If one considers the simulated data points with their statistical errors, there is however a statistically significant signal above the background. Background subtraction can improve the visibility and is done here in two steps. First the non resonant background is fitted with a polynomial of degree eight and then subtracted. It is important to know the shape of this background. This information can be obtained from detailed Monte Carlo simulations or experimentally by varying the $b$-tagging quality. Figure~\ref{fig:wh_peaks} (left) shows the result of this subtraction: one can see a double peak which is a superposition of the Higgs boson and the $Z^0$ peaks. Even for a very small Higgs signal the $Z^0$ peak should be still visible. This is a good cross check of the subtraction method, and can be used to calibrate the Higgs mass. In a second step the $W^\pm Z^0$ events are subtracted. The accurate magnitude of this background can be estimated easily from the analysis of $W^\pm Z^0 \rightarrow l^\pm \nu l^+ l^-$ events. The result is shown in Figure~\ref{fig:wh_peaks} (right) and the Gaussian fit of the Higgs peak is in good agreement with the simulated pure signal and the Higgs mass can be fitted.

With this type of analysis we obtain following results for $m_{H^0} =$ 115 $GeV/c^2$ and $L_{int} = $ 300 $fb^{-1}$: signal to background ratio $S / B =$ 2.3\% , significance $S / \sqrt{B} =$ 6.0 , precision on $WWH$ coupling $\Delta g_{WWH} / g_{WWH} =$ 8.4\% and precision on the mass $\Delta m / m =$ 2.3\%. $S$ and $B$ are the number of events in the mass window around the Higgs mass peak. The precision on the mass is obtained from the Gaussian fit in Figure~\ref{fig:wh_peaks} (right).


\section{Expectations for SM Higgs}

\begin{figure}[ht]
\begin{center}
 \includegraphics[width=0.58\textwidth,angle=+0]{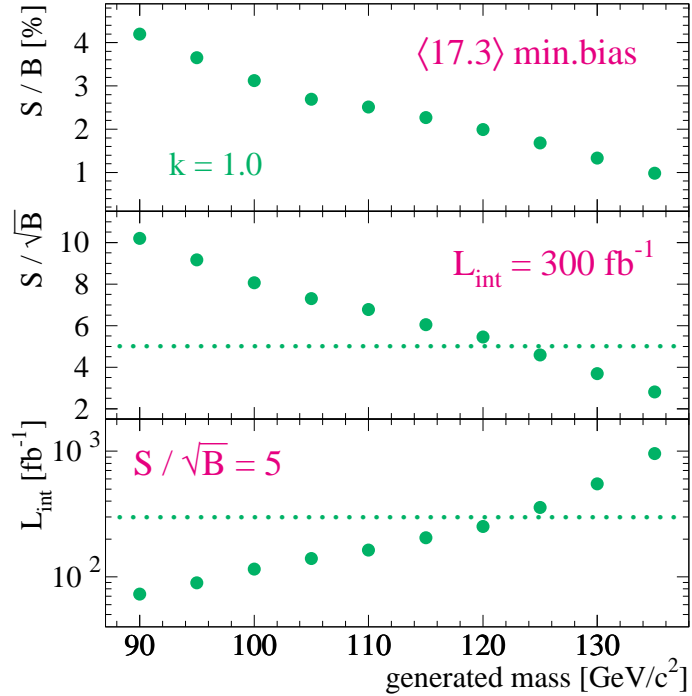}
 \caption{\sl $S / B$, $S / \sqrt{B}$ ($L_{int} = $ 300 $fb^{-1}$) and $L_{int}$ required for $S / \sqrt{B} =$ 5 versus generated SM Higgs mass in the $W^\pm H^0 \rightarrow l^\pm \nu b\bar{b}$ channel. No k-factors (k = 1.0) are used.\rm}
 \label{fig:wh_stat}
\end{center}
\end{figure}

\begin{figure}[ht]
\begin{center}
 \includegraphics[width=0.58\textwidth,angle=+0]{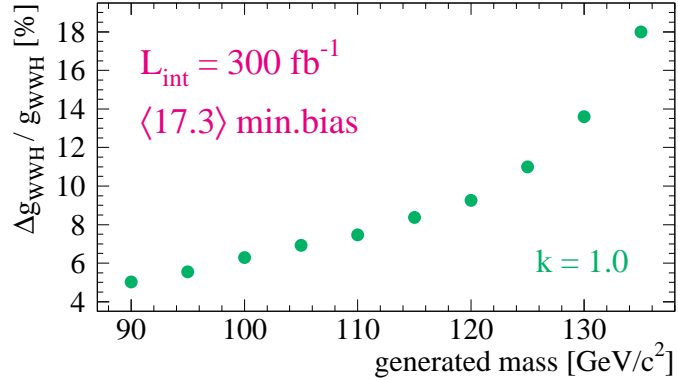}
\vspace*{-2mm}
 \caption{\sl Precision of $WWH$ coupling as a function of generated Higgs mass in the SM. $L_{int} = $ 300 $fb^{-1}$ is assumed and no k-factors are applied.\rm}
 \label{fig:g_wwh}
\end{center}
\end{figure}

The SM results are shown in Figures~\ref{fig:wh_stat} and \ref{fig:g_wwh}. For $L_{int} = $ 300 $fb^{-1}$ the 5 $\sigma$ discovery limit is at $m_{H^0_{SM}} \le$ 123 $GeV/c^2$. A discovery during the low luminosity phase ($L_{int} < $ 30 $fb^{-1}$) is not expected in this channel - even in the easiest cases i.e. at low Higgs masses already excluded by LEP. Higgs boson discovery is expected in another channel. The $W^\pm H^0$ channel is nonetheless interesting, 
\vspace*{5mm}
because it allows the measurement of the Higgs couplings which will tell us something about the nature of this particle. The $WWH$ coupling is proportional to the square root of the number of signal events. The expected precision in the measurement of the $WWH$ coupling is given in Figure~\ref{fig:g_wwh}, and is of the order of 10\% assuming a known branching ratio $BR(H^0 \rightarrow b\bar{b})$.
\vspace*{3mm}

Even if it is impossible to determine $BR(H^0 \rightarrow b\bar{b})$ in absolute value, the different signals of various Higgs production mechanisms can be exploited to determine ratios of Higgs couplings. For example, the ratio of $g_{WWH}$ and $y_t$ (top Higgs Yukawa coupling constant) is independent of $BR(H^0 \rightarrow b\bar{b})$ and can be determined with the $t\bar{t}H^0$ and $W^\pm H^0$ channels. The ratio of $BR(H^0 \rightarrow b\bar{b})$ and $BR(H^0 \rightarrow \gamma \gamma)$ can be determined by comparing $W^\pm H^0$, $H^0 \rightarrow b\bar{b}$ with $W^\pm H^0$, $H^0 \rightarrow \gamma \gamma$ final states produced by the same $q\bar{q}\prime \rightarrow W^* \rightarrow W^\pm H^0$ mechanism.


\section{MSSM Results}

To give an idea about the discovery potential of the corresponding channel $W^\pm h^0 \rightarrow l^\pm \nu b\bar{b}$ in the MSSM, we extrapolate the SM results (by rescaling the production cross section times branching ratio, obtained with SPYTHIA~\cite{SPYTHIA}) and discuss the parameter space coverage of one benchmark scenario called maximum $m_h$ scenario \cite{SUSYbenchmark}. This scenario turns out to be the most unfavourable one as $h^0 \rightarrow b\bar{b}$ visibility decreases with increasing $h^0$ mass. The reason is the rapidly falling cross section and branching ratio with increasing Higgs mass which limits the discovery potential of this channel in the SM as well.
\vspace*{3mm}
\begin{figure}[ht]
\begin{center}
 \includegraphics[width=0.49\textwidth,angle=+0]{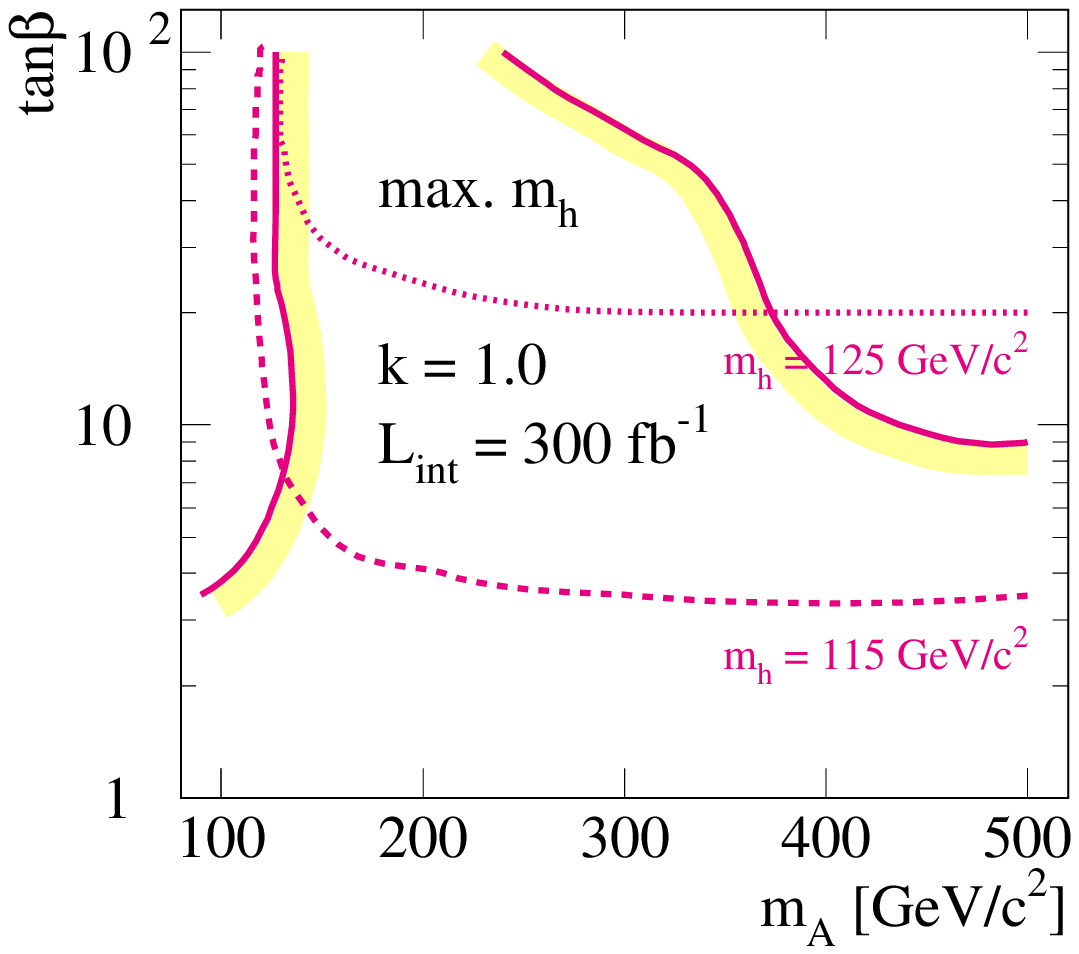}
 \includegraphics[width=0.49\textwidth,angle=+0]{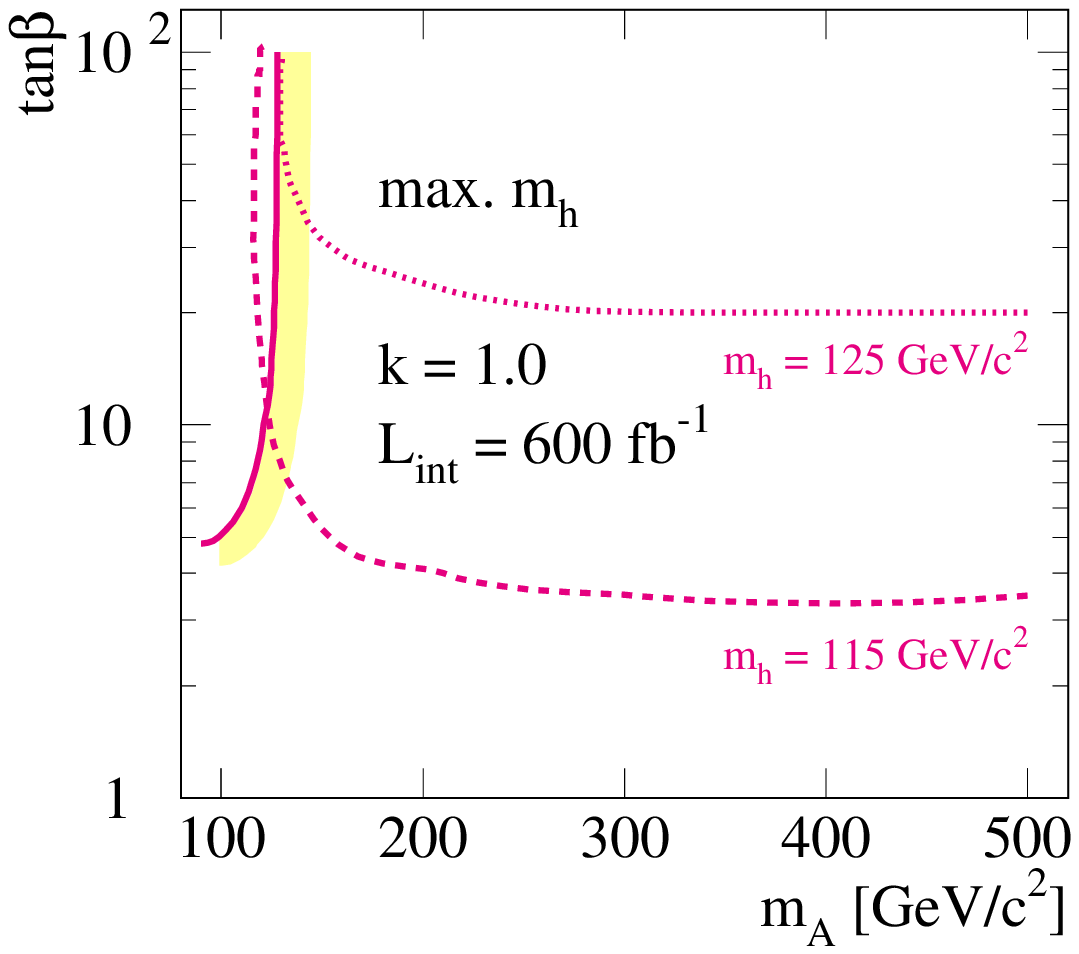}
\vspace*{-2mm}
 \caption{\sl Discovery contours in the MSSM (maximum $m_h$ scenario) parameter space for $L_{int} = $ 300 $fb^{-1}$ (left) and for $L_{int} = $ 600 $fb^{-1}$ (right). $S / \sqrt{B} \ge$ 5 to the shaded side of the solid line. The dotted and dashed lines are the isomass curves for $m_{h^0} =$ 125 $GeV/c^2$ and $m_{h^0} =$ 115 $GeV/c^2$, respectively.\rm}
 \label{fig:wh_susy}
\end{center}
\end{figure}

Figure~\ref{fig:wh_susy} shows the parameter space coverage in the $m_A$  $\tan\beta$ plane for two integrated luminosities integrated over very long LHC running periods (5 - 10 years). In both cases there is an inaccessible region at low $m_A$; the second difficult region at high $m_A$  and $\tan\beta$ disappears with increased integrated luminosity. In scenarios with non maximum $m_h$ the difficult regions are somewhat reduced implying that for high enough integrated luminosity most of the MSSM parameter space can be explored with this channel.


\section{Conclusions}

At this level of simulation, including high luminosity event pile up, we conclude that the selection and reconstruction of the $W^\pm H^0 \rightarrow l^\pm \nu b\bar{b}$ signal is possible and provides useful information. Excellent $b$-tagging performance and good mass resolution is crucial for a successful analysis. In addition, track and primary vertex reconstruction is important for the separation of jets from the hard interaction from jets of pile up events.

A very high integrated luminosity of 300 $fb^{-1}$ is necessary in order to have reasonable sensitivity to this signal: in the SM the discovery is limited at a Higgs mass of $<$ 123 $GeV/c^2$. In the MSSM a large fraction of the parameter space can be covered. It is clear that the very large integrated luminosity required makes this channel inappropriate for a first discovery of the Higgs boson, in contrast to the channel $t\bar{t} H^0$, $H^0 \rightarrow b\bar{b}$ \cite{TTH}.

The $W^\pm H^0 \rightarrow l^\pm \nu b\bar{b}$ channel allows direct measurement of the $WWH$ coupling for known branching ratio $BR(H^0 \rightarrow b\bar{b})$. Even if $BR(H^0 \rightarrow b\bar{b})$ is not known, this channel can be combined with other channels with $H^0 \rightarrow b\bar{b}$ decay \cite{THESIS} and ratios of the corresponding coupling strengths can be determined which would help to better understand the nature of the Higgs boson.


%
\end{document}